# Fostering Self-Directed Growth with Generative AI: Toward a New Learning Analytics Framework

Qianrun Mao

*Abstract*

In an era increasingly shaped by decentralized knowledge ecosystems and pervasive AI technologies, fostering sustainable learner agency has become a critical educational imperative. This paper introduces a novel conceptual framework integrating Generative Artificial Intelligence (GAI) and Learning Analytics (LA) to cultivate Self-Directed Growth—a dynamic competency enabling learners to iteratively drive their own developmental pathways across diverse contexts. Building upon critical gaps in current Self-Directed Learning (SDL) and AI-mediated educational research, the proposed Aspire to Potentials for Learners (A2PL) model reconceptualizes the interplay of learner aspirations, complex thinking, and summative self-assessment within GAI-supported environments. Methodological implications for future intervention designs and data analytics are discussed, positioning Self-Directed Growth as a pivotal axis for designing equitable, adaptive, and sustainable learning systems in the digital era.

## 1. Introduction

The educational realm faces two increasingly prominent challenges that threaten to reshape the landscape of learning and development. Firstly, the traditional teacher-dominated, institution-centered environment is being eclipsed by a decentralized, ever-evolving, and technologically advanced online landscape. In this new paradigm, knowledge and skills are not poised and delivered by a single expositor, but are constantly renewed, reproduced, and reiterated through sharing and co-creation, rendering existing models of education insufficient.
And the overreliance on EdTech tools, as well as information search and synthesis tools, such as Generative Artificial Intelligence (GAI), among students poses a significant challenge in the contemporary educational landscape, while there is a concerning lack of research examining whether these tools genuinely foster the development of learner agency.

The integration of AI into educational practices offers a transformative opportunity to enhance learning outcomes and promote equity. According to the United Nations Educational, Scientific and Cultural Organization (UNESCO), AI has the potential to accelerate the achievement of Sustainable Development Goal 4 (SDG 4) by improving access to quality education for all learners, regardless of their socioeconomic background (UNESCO, 2019; UNESCO, 2021).

As some noted, AI facilitates access to information and online education, helping to bridge the information, skill, and educational gaps faced by disadvantaged individuals who encounter barriers to traditional learning opportunities due to time constraints, financial limitations, geographic distance, or physical challenges (Thakkar et al., 2020; Sanabria-Z et al., 2023). ChatGPT enhances student productivity and supports academic achievement by providing valuable information and resources (Fauzi et al., 2023).

With the implementation of an omniscient digital infrastructure and AI support, it seems evident that all students could potentially access high-quality, adaptive learning experiences and significantly advancing educational equity. However, it is essential to understand that access barriers are not the sole contributors to educational gaps. In today's information age, students navigating the digital landscape often find themselves inundated with a torrent of information. While access to knowledge has improved in many countries and regions, pervasive issues of educational homogenization and standardization persist, severely impacting educational equity. Such an approach restricts students' capacity to envision and pursue diverse pathways aligned with their unique experiences and aesthetic preferences, thus hindering equal educational opportunities (Appadurai, 2004; Zipin et al., 2013; Gale, 2014).

The concept Capacity to Aspire is proposed by Appadurai (2004), who argues that disadvantaged populations are often less able to narrate the relationships between "Specific goods and outcomes, often material and proximate," and the "pathways through which bundles of goods and services are actually tied to wider social scenes and contexts, and to still more abstract norms and beliefs"; and the capacity to "navigate the cultural map in which aspirations are located and to cultivate an explicit understanding of the links between specific wants or goals and more inclusive scenarios, contexts and norms" is the essentiality of a strong Capacity to Aspire.

Similarly, Bauman (2009), pointed out in his delineation of "liquid modernity", the very process of unleashing and expanding the 'inner forces' presumably hidden within one's personality, waiting to be awakened and put into action, embodies the type of knowledge—more accurately described as inspiration—that individual seeks through education. Indeed, what truly matters for equitable education is Capacity to Aspire, the understanding, recognition, and envisioning of increasingly well-defined developmental pathways, where visions and aspirations guide and uphold concrete practices in pursuit of learners' envisioned futures, alongside the choices they make and the knowledge, skills, and strategies they develop.

This perspective is further reinforced by the Pearson Global Learner Survey (2019), which surveyed over 11,000 participants across 19 countries. The findings revealed an urgent need for lifelong, self-directed learning to address the demands of academic and professional development, particularly during transitions between educational and career phases.

As United Nations Educational, Scientific, and Cultural Organization (UNESCO, 2019) has highlighted, the application of AI in education faces significant challenges, necessitating increased human capital training to identify and implement the most effective educational uses of GAI. In this nascent field, it is crucial to prioritize the potential of educational equity, not merely through standardization, but by building self-directed competencies, that serve as a guiding compass to empower individuals to achieve their diverse personal goals, driven by their own initiatives in a continuous, sustainable learning and developmental process.

This study aims to strengthen educational equity by enabling learners to conceptualize clear developmental pathways, where visions and aspirations act as guiding principles and foundations for concrete self-directed learning practices. It also addresses a critical research gap by integrating the foundational pillars of self-directed learning into a novel conceptual framework. This framework incorporates learning analytics to foster learner agency and support the development of self-directed learning competencies across diverse tasks and contexts through continuous and iterative cycles of learning and growth.

## 2. Literature Review

**Challenges in Integrating Generative AI with Self-Directed Learning: A Critical Review**

To empower individuals in building the personalized and strategic pathway that serves as a guiding compass for planning and continuous and autonomous learning, the education system requires a new structural approach that diverges from the traditional, rigid teacher-centered model — an approach that encourages personalized learning approaches within a guided learning process that focuses on strengthening learner agency. Learner agency refers to the capacity of learners to take initiative and ownership of their educational journey, making choices and decisions that shape their learning experiences and outcomes (Bandura, 2006). One of the foundational student-centered approaches here is self-directed learning.

Self-directed learning (SDL) was first proposed by various researchers (Knowles 1970, 1975; Rogers 1969; Tough 1971) to create a learner-centered learning experience, which positions the learner as actively managing their educational journey, encompassing the identification of learning needs, goal setting, resource selection, strategy implementation, and outcome evaluation, which essentially fosters learner agency.

Recent literature highlights the potential of GAI to enhance SDL. For instance, researchers have explored how AI-driven systems can support SDL by providing adaptive learning resources, enabling personalized feedback, and facilitating skill acquisition based on individual goals and progress (Zawacki-Richter et al., 2019; Chassignol et al., 2018). GAI also evinces great potential in serving as an enabler of personalized education, providing continuous feedback and alignment with labor market demands and remains competitive in a knowledge-based economy (Troka, 2022; Chen & Liu, 2018; Pérez-Ortiz et al., 2020; Ally & Perris, 2022; Tang & Deng, 2022; Sanabria-Z et al., 2023).

Yet, two issues remain. Firstly, how to design an innovative SDL framework that addresses the broader developmental goal of cultivating a spiral structure of continuous, upward-learning progression—one that transitions across different contexts and tasks and, more importantly, evolves toward a person-oriented focus, namely the capacity to aspire?

Secondly, given GAI's analytical prowess and its adaptability for personalized feedback, what adaptation can be employed to leverage GAI for the effective collection of learning analytics to foster guided discovery through structured, real-time feedback, thereby supporting learners' SDL processes instead of merely adapting to learners' needs and reinforcing over-reliance?

There is an emerging research focus on this area, with various studies highlighting the advantages and disadvantages of GAI in personalized learning, though not necessarily facilitating SDL (Kasneci et al., 2023; Zou et al., 2023). A significant body of research examines self-directed language learning in lack of intervention (Lashari & Umrani, 2023), while rising concerns regarding ChatGPT focus on its impact on privacy violations (Dempere et al., 2023), students' academic integrity (Cotton et al., 2023; Rudolph et al., 2023), and users' over-reliance on GAI (Gao et al., 2022; Kasneci et al., 2023). Roe and Perkins (2024) identified 24 articles published between 2020 and 2024 that delve into the intersection of GAI and SDL in greater depth. Although a substantial body of research evaluates learners' SDL competencies (Wu et al., 2024; Indriani et al., 2024; Han et al., 2022, among others), the majority of these studies do not prioritize the structural and guided cultivation of SDL skills for learners.

Shalong et al. (2024) and Wang et al. (2024) present studies that are most closely aligned with the current research focus. Wang et al. (2024) primarily employs a teacher-led, GAI-assisted learning approach to demonstrate GAI's effectiveness in promoting SDL competencies. However, this study does not adequately distinguish between SDL and Self-Regulated Learning (SRL) within its research design. Instead, it emphasizes SRL indicators such as self-efficacy, task-related learning motivation (as opposed to the Capacity to Aspire), and technology acceptance to represent learners' growth in SDL.

In contrast, Shalong et al. (2024) developed LearnGuide, a customized version of ChatGPT specifically designed to enhance SDL skills. LearnGuide provides real-time, personalized feedback encompassing SDL strategies and solutions, including self-assessment, goal setting, active learning, and evaluation for user-generated prompts. This may potentially foster an over-reliance on LearnGuide.

Furthermore, both studies predominantly rely on post-task questionnaire data to determine whether participants have acquired specific SDL competencies, such as planning, within pre-assigned tasks set by the researchers. In the absence of support from learning analytics, this methodology assumes that students have already internalized these competencies. Consequently, the participants' responses may merely reflect superficial engagement or increased familiarity with procedure-related terminology acquired from the GAI, such as LearnGuide. This significantly limits the objectivity and adequacy of these studies in evaluating the sustainability and transferability of SDL skills over a continuous cycle on a personal level. Such an approach does not effectively capture the developmental process necessary for facilitating upward progression that transitions seamlessly across different tasks and contexts.

Considering that self-regulated skills are frequently assessed within the domain of SRL studies and given the increasing number of AI-focused studies in this area, it is worth exploring whether these fields might offer valuable insights for advancing our understanding and evaluation of SDL.

Emerging studies integrating AI (including GAI) with SRL have also predominantly focused on structured classroom settings and pre-designed curricula, similarly, they primarily used questionnaire-based data collection to evaluate self-regulated skills (Chang, Lin, Hajian, & Wang, 2023; Jin et al., 2023; Hsu, Chang, & Jen, 2023; Chiu, 2024; Ng, Tan, & Leung, 2024).

And among them, some have noticed the importance of guided SDL application within the administrated task, however, there are other serious issues remain. Ng, Tan, and Leung (2024) compared students using SRLbot, a ChatGPT-enhanced bot, with those using the rule-based Nemobot. SRLbot provided real-time, adaptive, personalized solutions for knowledge-based science learning as well as self-regulated learning (SRL) strategies.

While from post-task survey data and backend system data, it stated that SRLbot enhances students' SRL as it effectively increases task motivation and reduces anxiety, the study raises concerns that its use may overlook learner agency and obscure the complex interplay of decision-making, motivation, and self-regulation between the GAI and learners (Turner & Patrick, 2008). As several learners in experiment group demonstrated over-reliance, for example, one of them expressed: "SRLbot helps me set clear goals, create schedules and routines, and organize my study materials."

Similar to Shalong et al. (2024), this problematic outcome arises because this research design positions task-specific outcomes and task-related self-efficacy as the indicators or manifestations of SRL improvement, which are essentially facilitated by the GAI. Similar to aforementioned SDL studies, learning behavior observed within a single task merely serves as a demonstration of engagement. It is the sustained and purposeful engagement in learning behaviors, driven by the individual's intentional application of strategies across multiple contexts, that can be framed as the indicators of the development of competencies, contributing to an upward trajectory of growth, to the core of SDL.

Thus, the crux of the issue, as repeatedly highlighted in prior research, is that we cannot rely on task-level engagement of "learner agency" to assess learner agency which is essentially a self-initiated and continuous learning process on person level, as doing so risks circular reasoning, oversimplifies its complexity, and neglects critical dimensions such as context, strategies, and feedback integration in the use of GAI. There remains an overarching need for an adapted SDL framework that integrates open pedagogical design and robust learning analytics design.

**The technical possibility of Leveraging Learning Analytics to focus on summative self-assessment in GAI**

To effectively facilitate self-directed growth, defined as a learner's capacity to drive continuous and sustainable self-directed learning (SDL) cycles across various contexts and tasks, GAI needs to intentionally collect learning data traces that reflect learning progress and provide targeted feedback.

Learning Analytics (LA) refers to the practice of collecting, measuring, and analyzing data about learners and their contexts to understand and optimize learning processes and environments (Siemens & Baker, 2012). It leverages data analytics techniques to gain insights into how students learn, identify challenges, and improve educational outcomes (Ferguson, 2012; Papamitsiou & Economides, 2014). By analyzing educational data, educators can personalize learning experiences and enhance educational interventions (Long & Siemens, 2011; Clow, 2013)

While GAI holds significant potential, practical studies exploring possibilities of integration of GAI and LA remain scarce. Yet, Chui's (2023) assertion that GAI technologies assess students' (users') performance is grounded only in conceptual frameworks also lacks validity upon closer scrutiny. In fact, GAI systems, such as ChatGPT, are designed for tasks like formative assessment in response to user inquiries. Built on a deep learning architecture, these systems rely on a Transformer-style model pre-trained to predict the next token in a sequence, with additional fine-tuning through reinforcement learning from human feedback (Ouyang et al., 2022; OpenAI et al., 2024). GAI's capability to process multimodal inputs (e.g., text and images) and generate text outputs that are "better aligned with the user's intent" (OpenAI et al., 2024) is essentially its own adaptive learning process. This functionality inherently encourages over-reliance on the system (Passi & Vorvoreanu, 2022; Zhai et al., 2024) rather than fostering the collection of meaningful, learner-centric data that is critical for SDL development or guiding summative self-assessment. Failure to address this issue risks perpetuating significant research gaps, as outlined in the discussion of existing research.

Despite the scarcity of research that directly combines ChatGPT and learning analytics, this section reviews studies that collectively provide a foundational basis and are structurally mapped out according to the research design.

According to Yan et al. (2024), Generative Artificial Intelligence (GAI) can be contextualized within the learning analytics loop by integrating four interconnected components: learners, data, analytics, and interventions (see Figure 1). In this research, the learner component includes students and their interactions with AI agents, while the data extracted consists of summative self-assessment frameworks across learning cycles. By employing both diagnostic and interactive analytics, this study bridges two critical stages of learning analytics: insight generation and feedback delivery.

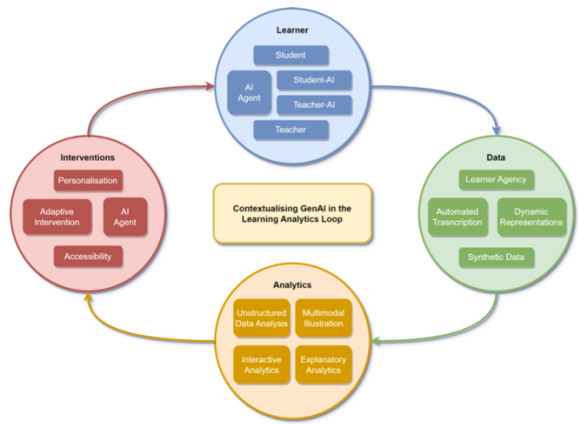

Figure 1. Contextualization of generative artificial intelligence within Clow's Learning Analytics Cycle. Adapted from Yan, at. el (2024)

Diagnostic analytics serve as the foundation by identifying patterns and trends in how learners reflect on and adjust their self-assessment framework over time. Building on this, interactive analytics foster dynamic engagement by generating guiding questions tailored to the learners' submitted assessments. These interactions encourage iterative refinement and deeper reflection, while carefully maintaining a non-prescriptive approach that avoids revealing GAI's evaluation or offering direct solutions.

In the diagnostic approach, summative self-assessment is essentially a complex thinking task. Sanabria-Z et al. (2023) designed an AI-based platform to assess students' complex thinking through an ideathon task centered on the Shared Economy. This framework employs decision tree algorithms and a relational database to assess complex thinking traits by systematically collecting and analyzing user activity data. The decision tree validates responses and classifies elements based on a rubric that defines mastery levels for complex thinking. The system segments data into homogeneous sub-regions to ensure precise classification, with results stored in a relational database for further analysis. By combining automated evaluation with a robust rubric, the framework provides insights of how to assess students' summative self-assessment in GAI.

As for interactive analytics, the key lies in providing real-time feedback that is meaningful yet non-prescriptive. Instead of offering direct answers or evaluation outcomes, GAI can pose guiding questions that prompt deeper reflection and encourage learners to independently build the summative self-assessment frameworks. In this research, GAI acts as a facilitator, an assessor, but essentially a collaborator, the interaction between GAI and students can be conceptualized as a collaborative knowledge-building process within a guided discovery process. Scardamalia and Bereiter's (2006) work on knowledge building emphasizes the iterative refinement of ideas through collaborative scaffolding, which is echoed in the Spiral Model of Collaborative Knowledge Improvement (Chen et al., 2019). This research has explored structured approaches to learning analytics, often implementing staged, task-specific data extraction to monitor and support learning progression and using phase-based data tracking and feedback loops within group settings to foster collaborative knowledge refinement (Chen et al., 2019). This study provides insights into structuring learning analytics that develop and reinforce the pedagogical goal through stages where collaborative learning process is fostered.

Several studies have examined the adaptation of ChatGPT's role as an assistant, with a particular emphasis on refining its ability to formulate text outputs in a way that fosters active engagement. Among these, Ali et al. (2023) introduced TeacherGAIA, a framework designed not to provide direct answers, but to offer personalized guidance that promotes deeper engagement within K-12 learning environments. Although it does not assess learners' SDL competency, this work highlights the potential of generative AI models like GPT-4, which, with dynamic prompts, can rapidly perform a diverse array of tasks without requiring additional fine-tuning. The emergent property of in-context learning exemplifies the versatility of prompt engineering, suggesting a transformative potential for creating interactive, personalized learning experiences.

## 3. Conceptual Framework: Aspire to Potentials for Learners (A2PL)

This paper introduces the concept of Self-Directed Growth: it is a competency reflects by the increased progression to drive an ongoing, iterative, and sustainable SDL process independently, fueled by an enhanced Capacity to Aspire that underpins self-development and self-actualization. Self-Directed Growth highlights learner agency by positioning the learner as the initiator, executor, assessor, and primary agent responsible for their own SDL journey, in itself and by itself a dynamic and developmental process for the learner. In short, self-directed growth is not merely about the SDL process aimed at achieving predefined learning outcomes through pre-assigned tasks within a structured curriculum or context; rather, it represents the broader ability to independently and continuously construct and assess one's own SDL process, guided by an internal compass. Self-Directed Growth can be considered an advanced and comprehensive extension of the broader SDL framework, representing its most developed and integrative structure.

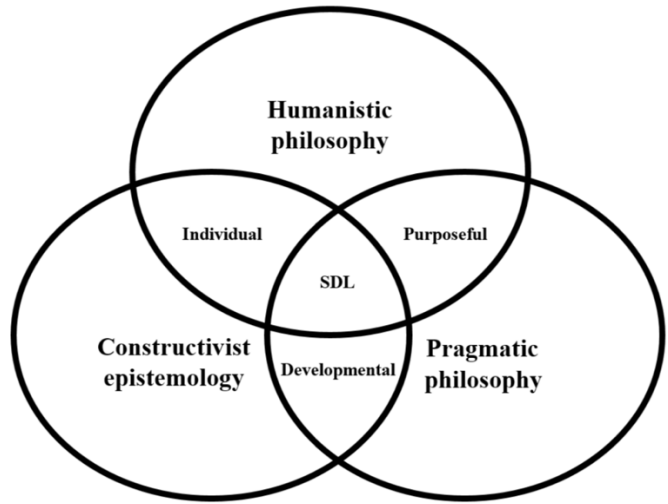

Figure 2. The Foundational positions of self-directed learning (SDL), Adapted from Morris (2019)

This paper reconfigures the three foundational pillars of Self-Directed Learning (SDL) as summarized by Morris (2019) (see Figure 2) to build the framework of Self-Directed Growth. The Capacity to Aspire represents the core essence of humanistic philosophy; Complex Thinking serves as the foundation of constructivist epistemology; and Self-Assessment embodies the principles of pragmatic philosophy. At the center of this framework is Self-Directed Growth, which, together with the other three pillars, represents a learning process that is individual, purposeful, and developmental (see Figure 3). The following sections will discuss each of these three foundations in detail.

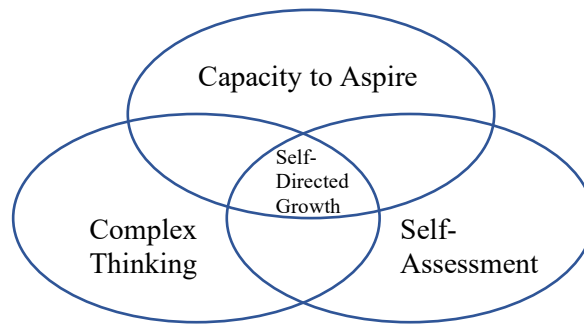

Figure 3. The Framework of Self-directed Growth

**The humanistic perspective: Capacity to Aspire**

The humanistic perspective, regards learning goals as pathways for personal development, emphasizing growth as an ongoing, developmental process, which are inherently driven toward self-actualization and infinitive growth potential, shaped by each learner's self-concept and individual understanding of the world (Groen & Kawalilak, 2014; Elias & Merriam, 1995;

Leach, 2018). This aligns with Capacity to Aspire as the ability to "navigate the cultural map in which aspirations are located and to cultivate an explicit understanding of the links between specific wants or goals and more inclusive scenarios, contexts, and norms." (Appadurai, 2004) In the context of this research, Capacity to Aspire consists of two interconnected layers. The first involves identifying aspirations and the long-term vision of societal achievement and professional growth. The second layer encompasses a reflective mindset toward resources, skills, and strategies gaps, which aims to foster alignment between one's current state and envisioned future, where the developmental process is actively engaged and fully actualized.

**Constructivist epistemology: Complex Thinking**

Constructivist epistemology emphasizes active knowledge construction, where knowledge is subjective, personal, and developed through interactions with others and authentic real-world situations (Cobb & Bowers, 1999; Simpson, 2002; Schunk, 2020; Jonassen, 1999). It aims to cultivate learners to approach complex issues with a reflective mindset (Brookfield, 1985). Within this framework, Complex Thinking is particularly suited for active knowledge construction, as it involves developing advanced analytical and adaptive reasoning abilities that empower learners to address multifaceted, real-world challenges (Schunk, 2020).

Furthermore, studies affirm the established correlations between metacognition and critical thinking (Barzegarbafrouee, Farzin, & Zare, 2019; Amin, Corebima, Zubaidah, & Mahanal, 2020; Arslan, 2018; Correa, Ossa, & Sanhueza, 2019; Lukitasari, Hasan, & Murtafiah, 2019), a component of complex thinking (Ramírez-Montoya et al., 2022). And Silva (2020) also highlights the alignment of complex thinking tasks with the goal of developing learners' capacity to "…critically or self-critically examine their own procedures and methodologies, thereby self-regulating their thought processes (metacognition)". Thus, complex thinking tasks not only serve as a context for cultivating advanced analytical and adaptive reasoning skills applicable to real-world scenarios but also function as essential training for fostering the reflective mindset necessary for self-assessment in subsequent learning stages.

**The pragmatic philosophy: Self-assessment**

The pragmatic philosophy focuses on the effectiveness of self-directed learning as a positive reinforcement loop where key skills are produced and reproduced, emphasizing learners' management and control over both the methods and goals of their learning (Brookfield, 1986; Gibbons, 2002; Grow, 1991; Mocker & Spear, 1982). Self-assessment can be an effective tool for students to monitor and manage their own learning (Pintrich, 2000; Zimmerman & Schunk, 2001), enabling students to generate their own feedback, promote continuous learning, and enhance performance (Andrade, 2019).

As suggested by Andrade (2019), Self-assessment encompasses a broader range of processes and outcomes, including both formative and summative evaluations of competence, processes, and products. Formative assessment revolves around self-efficacy ratings of a specific task, summative self-assessment emphasizes post-task evaluations, including judgments of ability based on performance and procedures. This approach moves beyond immediate self-efficacy ratings to a broader analysis of the effectiveness and outcomes of learning processes.

Previous research in this area has predominantly employed a mixed approach. While self-efficacy assessments may be collected and analyzed by backend data, summative self-assessments are frequently conducted by researchers as post-task questionnaires in evaluations of students' SRL skills based on demonstrated performance in previous essays, which, as previously discussed, is ineffective. This research adopts an alternative approach to assessing students' self-assessment skills, with a specific focus on learner-driven summative self-assessment. Yet, summative self-assessment alone cannot drive an ongoing, iterative, and sustainable SDL process, unless it is combined with Capacity to Aspire. This framework highlights the alignment of summative self-assessment with the two layers of Capacity to Aspire within and across learning cycles, with a central focus on the cultivation of Self-directed growth.

## 4. Methodological and Practical Implications

**The Aspire to Potentials for Learners framework as an innovative intervention design**

The Aspire to Potentials for Learners (A2PL) framework proposed in this study serves as an innovative intervention designed to address the gaps identified in the previous sections A2PL provides a structured approach to learning that empowers learners to be the initiators, assessors, and primary agents, who actively shape and take responsibility for their own strategic developmental pathways. This self-paced learning intervention emphasizes learner autonomy in managing the timing and progression of their involvement in a non-synchronous online learning environment. This intervention takes place in three levels for learners (see Figure 4):

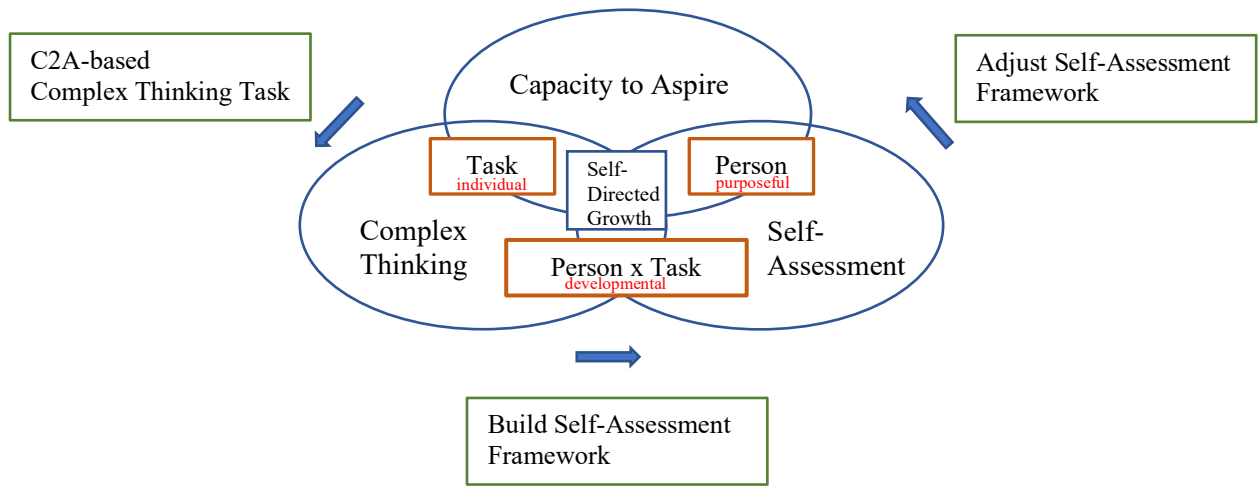

Figure 4. The framework of Aspire to Potential Learning Approach(A2PL)

**At the Person level**, the intervention prioritizes practices that are grounded in humanistic philosophy which are purposeful to the learners. In specific, learners will be guided to reflect on their aspirations, and the long-term vision of their societal achievement and professional growth.

Starting from the second round, it will integrate reflective practices to evaluate and adjust this alignment and continue to build on the summative self-assessment framework, while this time transcends individual tasks and focus once again on the broader vision of societal achievement and professional growth, which enables the initiation of a new cycle of SDL and guides the design of the next C2A-based Complex Thinking Task.

**At the Task level**, the intervention emphasizes constructivist epistemology through an individualized approach. Learners and GAI co-create complex thinking tasks within real world scenarios that align with learners' aspirations and their specific goals within their initial understanding of their personalized pathway. These tasks help develop learner's complex thinking skill which will be foundational to develop summative self-assessment frameworks and facilitate self-directed growth.

**At the Person x Task level**, the intervention focusses on the pragmatic philosophy where learners engage in the developmental process of an increasingly clear mapping of the pathway that drives learners to assess alignment of the acquired the resources, skills, and strategies with their potential. Learners are guided in constructing a summative self-assessment framework, achieved through the alignment of their reflective understanding of their abilities, skills, and strategy gaps identified during the task with their task-aligned personal aims. This process aims to foster the development of a more clearly defined and strategically aligned learning pathway.

The overarching aim of this intervention, however, is not to evaluate the overall completion rate or increased intricate pattern of an advanced strategic learning pathway as reflected in the summative self-assessment frameworks, and then frame it an indicator of effective summative self-assessment skills that represents "increased" self-directed growth. The increasement itself is not a static and isolated element to be represented in other learning behaviors or outcomes, but a dynamic, collaborative and interactive process; in specific, the process of summative self-assessment, aimed at cultivating a more precisely defined strategic developmental pathway aligned with personal aspirations; is intrinsically interconnected and mutually reinforcing to the targeted process: the development of self-directed growth, along with the pertinent SDL competencies and self-regulated skills, which is essentially a set of continuous SDL cycles as a learning pathway toward various learning goals(see Table 1). Thus, the success of this intervention will be contingent upon whether learners' summative self-assessment framework and its evolving trajectory adequately reflect analytical cognition of SDL into the depth of its procedures and components, situated within a more refined and contextually relevant strategic approach to their personal learning processes.

| Theoretical Guidelines | Pedagogical Goal for Learners | GAI Features | Skills Developed for Complex Thinking |
|---|---|---|---|
| Constructivist Learning Theory<br><br>Social Constructivism | Reflect on past experiences, identify interests, motivation, and aspiration | Examples of similar career paths of successful figures. | Innovative Thinking |
|  |  |  |  |
| Critical Pedagogy  Transformative Education | An existing issue that needs to be addressed. Conduct a root cause analysis to identify the underlying problem | Support learners in developing their root-cause analysis skills | Systemic Thinking |
| Metacognition, Self-Regulated Learning, and Reflective Thinking | Build a framework for self-assessment | Support learners in developing their self-assessment skills | Scientific Thinking |
| Problem-Based Learning, Project-Based Learning, & Interdisciplinary Learning | Select three core abilities Finish tasks | Offer a range of core abilities. Create scenario-based tasks aligned with the identified core abilities | Critical Thinking; Problem-solving Skills |
| Formative, Portfolio, and Authentic Assessments | Accomplish self-evaluation Adjust self-assessment framework | Generate feedback based on previous assessment framework | Critical Thinking |

Table 1. The learning stages and cycles of A2PL

While GAI can provide learners with a vast array of choices of information—including those pertaining to SDL—it will not intentionally facilitate the mastery of structured SDL knowledge. The GAI's role will remain restricted to guide learners for self-discovery in a learning journey where assessment rubrics are not explicitly provided. Therefore, as long as GAI refrains from disclosing that the assessment pertains directly to SDL, the capacity of learners' summative self-assessment framework to indicate progress in identifying specific areas for further development in an individualized context, and its relevant branching out to reflect analytical cognition of SDL into the depth of its procedures and components within contexts, will serve as a key indicator of their initial mastery of these skills and their integration into an ongoing, iterative learning process. This progression signals the successful cultivation of critical abilities such as self-monitoring, self-regulation, self-reflection, and self-evaluation on Task, Task x Person, and most essentially Person level, collectively facilitating the emergent development of self-directed growth.

**Learning analytics and the adaptation of GAI**

To realize the interventional process outlined above—where learners gradually internalize structured SDL competencies through guided discovery rather than explicit instruction—it is necessary to move beyond traditional, content-centric survey models. Achieving this objective requires the adoption of dynamic, learner-driven learning analytics approaches that can actively capture and foster learners' analytical engagement with SDL processes and strategies within personalized and evolving learning trajectories.

Within the layer of diagnostic analytics, the GAI evaluates students' summative self-assessment frameworks based on a pre-designed scoring rubric Aspire to Potential Scoring Rubric (APSR), that incorporates component weights, relational logistics, and vectorized progression, all mapped both within and across the phases of the SDL cycles. Component relevance, relational logistics, and vectorized progression refer to the tailored assessment of the relative importance of each element, the interconnections between them, and the progression through key stages of A2PL, all of which are evaluated to ensure alignment with the learner's goals and continuous development within and across cycles. Specifically, GAI examines learners' analytical cognition of SDL into the depth of its procedures and components, situated within a more refined and contextually relevant strategic approach to their personal learning processes. Importantly, the GAI's scoring assessment is not disclosed to learners.

Through interactive analytics, which are based on previous scoring assessments, the GAI delivers continuous, constructive, personalized, and targeted feedback, aimed at promoting guided discovery for the progression of students' summative self-assessment frameworks in alignment with their personal goals and the long-term vision of their societal achievement and professional growth. Specifically, feedback will be structured as thought-provoking, sequenced prompts, calibrated to an optimal level of cognitive challenge, designed to facilitate guided discovery, and promote deep reflective engagement in learners.

## 5. Limitations and Future Directions

While this study proposes an integrative conceptual framework that reimagines Self-Directed Growth through the interaction of GAI and LA, several theoretical limitations must be acknowledged.

First, although the A2PL model emphasizes learner agency and continuous self-directed development, its foundations remain largely conceptual. The framework assumes that learners will actively engage with GAI-supported feedback systems in a reflective and strategic manner. However, the variability in individual learners' intrinsic motivation, metacognitive maturity, and socio-cultural positioning may significantly influence their capacity to actualize such pathways. These human factors, deeply embedded within personal and contextual realities, are not fully accounted for in the present model.

Second, the framework presumes a relatively harmonious collaboration between learners and GAI systems, positioning AI as a non-prescriptive guide rather than a directive authority. Yet, the psychological dynamics of human-AI interaction—including the risks of perceived authority, externalization of judgment, and potential diminishment of internal self-regulation mechanisms—remain under-theorized. Future research must critically investigate how learners internalize or resist AI-mediated scaffolding, particularly across diverse socio-cultural and educational contexts.

Third, the notion of Self-Directed Growth, while conceptually distinct from traditional SDL, requires deeper theoretical refinement. Specifically, clearer operational definitions and distinctions between iterative self-directed cycles and aspirational long-term trajectories must be elaborated. Without empirical grounding, there remains a risk of conflating sustained motivation with strategic skill development, or oversimplifying the complex interplay between aspirations, agency, and learning processes.

Finally, this framework primarily draws upon humanistic, constructivist, and pragmatic philosophical traditions. While these perspectives robustly support the emphasis on learner-centered growth, the integration of critical theories—particularly those that interrogate issues of power, access, and technological determinism—could enrich the model. Future work could explore how socio-economic inequalities, cultural narratives of aspiration, and systemic educational structures intersect with or challenge the development of Self-Directed Growth.

In light of these limitations, future research should prioritize theoretical elaboration, empirical validation across diverse learner populations, and the critical interrogation of the socio-cultural assumptions embedded within AI-mediated educational frameworks. By doing so, the field can move toward a more nuanced, equitable, and sustainable vision of learner agency in the digital era.

## 6. Conclusion

The paper seeks to explore the effective application of GAI to facilitate Self-Directed Growth, conceptualized as learners' capacity to drive continuous and sustainable SDL cycles across diverse contexts and tasks. Central to this process is the enhanced development of a strategically refined learning pathway, aligned with both learners' immediate personal aims and their long-term visions of societal achievement and professional growth.

The proposed interventional framework A2PL, distinguishes itself from existing SDL approaches by integrating three interconnected layers—Person Level, Person x Task Level, and Task Level—into an organic system. This system cohesively fosters learners' analytical cognition of SDL procedures and components, situated within increasingly personalized and contextually relevant strategic pathways. By framing GAI not merely as a content provider but as a collaborative scaffold within this developmental architecture, A2PL seeks to cultivate learners capable of navigating the complexities of a rapidly evolving knowledge landscape with greater autonomy, resilience, and purpose.

This conceptual framework advances current discussions surrounding SDL, learning analytics, and AI in education by centering the learner's aspirational agency and iterative self-construction, rather than task completion or short-term performance gains. It emphasizes that sustainable educational equity in the digital age requires not only access to information and resources, but also the internalization of strategic capacities to envision, plan, assess, and refine personal developmental trajectories.

And it is important to clarify that the present study is conceptual in nature. The primary aim is to construct a theoretical framework that synthesizes key philosophical foundations, redefines core constructs, and proposes a coherent model for fostering self-directed growth in AI-mediated learning environments. Detailed technical implementations, platform engineering, or algorithmic design considerations fall outside the scope of this paper and are reserved for future research directions focused on empirical validation and application development.

Future research may continue to elaborate, empirically examine, and critically refine the principles of Self-Directed Growth proposed herein. As education increasingly intersects with AI-mediated environments, it becomes ever more crucial to ensure that technological interventions amplify, rather than diminish, learners' ownership of their futures. This study offers a conceptual steppingstone toward that vision.

**Acknowledgements**